\documentclass[preprint2]{aastex}

\newcommand{\vdag}{(v)^\dagger}

\slugcomment{Submitted to ApJ, May 7, 2019; 
Accepted on June 19, 2019}

\shorttitle{Electron-capture Rates for a Forbidden Transition in $^{20}$Ne}
\shortauthors{Suzuki, Zha, Leung, \& Nomoto}

\begin{document}


\title{Electron-capture 	Rates in $^{20}$Ne for a Forbidden Transition to 
the Ground State of $^{20}$F Relevant to the Final Evolution of High-density O$-$Ne$-$Mg Cores}


\author{Toshio Suzuki\altaffilmark{1}}

\affil{Department of Physics and Graduate School of Integrated
Basic Sciences,\\ 
College of Humanities and
Sciences, Nihon University\\
Sakurajosui 3-25-40, Setagaya-ku, Tokyo 156-8550, Japan}
\email{suzuki@phys.chs.nihon-u.ac.jp}

\author{Shuai Zha \altaffilmark{}}
\affil{Department of Physics, The Chinese University of Hong Kong, 
Shatin, N. T., Hong Kong S. A. R., China}


\author{Shing-Chi Leung and Ken'ichi Nomoto\altaffilmark{ }}
\affil{Kavli Institute for the Physics and Mathematics of the Universe (WPI),\\
The University of Tokyo, Kashiwa, Chiba 277-8583, Japan\\
\vspace*{0.5cm}
\rm{Submitted to ApJ, May 7, 2019; Accepted on June 19, 2019}}

\altaffiltext{1}{Visiting Researcher, National Astronomical Observatory of Japan, Mitaka, Tokyo 181-8588, Japan}



\begin{abstract}

Electron capture on $^{20}$Ne is critically important for the final stage of evolution of stars with the initial masses of 8 - 10 $M_{\odot}$.
In the present paper, we evaluate electron-capture rates for a forbidden transition 
$^{20}$Ne (0$_{g.s.}^{+}$) $\rightarrow$ 
$^{20}$F (2$_{g.s.}^{+}$) in stellar environments by the multipole expansion method with the use of shell-model Hamiltonians.
These rates have not been accurately determined in theory as well as in experiments.
Our newly evaluated rates are compared with those obtained by a prescription that treats the transition as an allowed Gamow-Teller transition with the strength determined from a recent $\beta$-decay experiment for $^{20}$F (2$_{g.s.}^{+}$) $\rightarrow$ $^{20}$Ne (0$_{g.s.}^{+}$) \citep{Kirsebom}.
We find that different electron energy dependence of the transition strengths between the two methods leads to sizable differences in the weak rates of the two methods.
We also find that the Coulomb effects, that is, the effects of screening on ions and electrons are nonnegligible. 
We apply our electron-capture rates on $^{20}$Ne to the calculation of the evolution of high-density O$-$Ne$-$Mg cores of 8 - 10 $M_{\odot}$ stars.  
We find that our new rates affect the abundance distribution and the central density at the final stage of evolution.
\end{abstract}





\section{Introduction}

The evolution and final fates of stars depend on their initial masses $M_{\rm I}$ \citep[e.g.,] {Nomoto13}, 
and are also subject to some uncertainties involved in stellar mass-loss, mixing processes, and nuclear transition rates.
A strongly electron-degenerate O$-$Ne$-$Mg core is formed after carbon burning in stars with $M_{\rm I} = 8-10 M_{\odot}$, which can end up in various ways, that is,
as O-Ne-Mg white dwarfs, or as electron-capture (e-capture) supernovae, or as Fe core-collapse supernovae \citep{Miyaji,Nomoto2,Nomoto3,Nomoto}.
The evolutionary changes in the central density and temperature of the degenerate O-Ne-Mg core are determined
by the competition among the contraction, cooling, and heating processes.

Nuclear URCA processes, especially in nuclear pairs with $A = 23$ and 25, are found to be important
for the cooling of the O$-$Ne$-$Mg cores after carbon burning \citep{Jones,TSN,Schwab2017,Schwab19}.
Electron-capture reactions and successive gamma emissions in nuclei with $A = 24$ and 20
are important for the contraction and heating of the core in later stages leading to 
an electron-capture supernova. 
The fate of the stars depends sensitively on the nuclear electron-capture and $\beta$-decay rates.
Accurate evaluations of the weak rates for high densities and temperatures in fine steps are important for a proper treatment of cooling and heating processes \citep{STN}. 

The weak rates for nuclei with $A =24$ and 20 are examined in detail in \citet{Pinedo14}
by taking into account the forbidden transitions between $^{20}$Ne (0$_{g.s.}^{+}$) and $^{20}$F (2$_{g.s.}^{+}$).
The forbidden transition was usually not taken into account to obtain the weak rates \citep{Taka}.
The forbidden transitions are found to give nonnegligible contributions at log$_{10}T <$ 9.0 in a density region; $9.3 < \log_{10}(\rho Y_e) < 9.6$. Here $Y_e$ is the proton fraction, namely, the lepton-to-baryon ratio.
The forbidden transitions, however, were treated as if they were allowed Gamow-Teller (GT) transitions and the $B$(GT) value was taken to be the largest one corresponding to the lower limit of the log~{\it ft} value of the $\beta$-decay: $ft = 6147 / B$(GT).
The experimental transition rate for the $\beta$-decay was not well determined: a lower limit of log~{\it ft} $>$ 10.5 is given in NNDC$^{5}$\footnotetext[5]{National Nuclear Data Center on-line retrieval system, http://www.nndc.bnl.gov}.

Recently, a new measurement on the $\beta$-decay has been carried out , and the transition rate is determined to be log~{\it ft} = 10.47 $\pm$ 0.11 \citep{Kirsebom}.
The mean value is very close to the lower limit value of log~{\it ft} = 10.5 (see footnote 5), 
and the difference is only by 7$\%$. 
However, in general for forbidden $\beta$-decays, shape factors are energy dependent and the prescription to use constant shape factors as in allowed transitions is an approximation. 

Here we treat the forbidden transitions between $^{20}$Ne (0$_{g.s.}^{+}$) and $^{20}$F (2$_{g.s.}^{+}$) properly and evaluate the weak rates by using the multipole expansion method \citep{Walecka}.
We compare the rates with those obtained by the prescription of using a constant $B$(GT) value assuming the transitions as allowed ones.
We also investigate the effects of the screening effects on the rates.  
The aim of the present paper is to point out the difference in the transition strengths and weak rates between the multipole expansion method and the prescription assuming allowed GT transitions.
We discuss origins and reasons that cause the differences.

In sect. 2, we discuss e-capture rates of the forbidden transition on $^{20}$Ne.
We also discuss $\beta$-decay rates of the forbidden transition from $^{20}$F$_{g.s.}^{+}$. 
In sect. 3, 
the dependence of the evolution of the O$-$Ne$-$Mg core on the e-capture rates is investigated in the later heating stages.  
Summary is given in sect. 4.

\section{Electron-capture rates on $^{20}$Ne}

We discuss e-capture rates for the forbidden transition, $^{20}$Ne (0$_{g.s.}^{+}$) $\rightarrow$ $^{20}$F (2$_{g.s.}^{+}$).
Formulae for the e-capture rate for finite density and temperature are given as \citep{Ocon, Walecka, Parr,Vretenar},
\begin{eqnarray}
\lambda^{ecap}(T) &=& \frac{V_{ud}^2 g_V^2 c}{\pi^2 (\hbar c)^3}
\int_{E_{th}}^{\infty} \sigma(E_e,T) E_e p_e c f(E_e) dE_e \nonumber\\
\sigma(E_e,T) &=& \sum_{i} \frac{(2J_i +1)e^{-E_i/kT}}{G(Z,A,T)}
 \sum_{f}\sigma_{f,i}(E_e)\nonumber\\
G(Z,A,T) &=& \sum_{i} (2J_i +1) e^{-E_i/kT}, 
\end{eqnarray}
where $V_{ud} = \cos \theta_C$ is the up-down element in the Cabibbo-Kobayashi-Maskawa quark mixing matrix with $\theta_C$ as the Cabibbo angle, $g_V = 1$ is the weak vector coupling constant, $E_e$ and $p_e$ are electron energy and momentum, respectively, $E_{th}$ is the threshold energy for the electron capture, and
$f(E_e)$ is the Fermi-Dirac distribution for electron.
The electron chemical potential is determined from $\rho Y_e$ with $\rho$ the baryon density and $Y_e$ is the proton fraction.
Here, $i$ denotes the initial state with excitation energy $E_{i}$ and angular momentum $J_i$, and $f$ specifies the final state.  
The cross section $\sigma_{f,i}(E_e)$ from an initial state with $E_i$ and $J_i$ to a final state with excitation energy $E_f$ and angular momentum $J_f$ is evaluated with the multipole expansion method \citep{Ocon,Walecka}. 
\begin{eqnarray}
\sigma_{f,i}(E_e) = \int (\frac{d\sigma}{d\Omega})_{f,i} d\Omega \nonumber\\
(\frac{d\sigma}{d\Omega})_{f,i} 
= \frac{G_F^2} {2\pi} \frac{F(Z,E_e)}{(2J_i+1)}
(\sum_{J\leq1}W(E_{\nu})\nonumber\\
\times \{(1-(\hat{\vec{\nu}}\cdot\hat{\vec{q}})(\vec{\beta}\cdot\hat{\vec{q}}))  
[|\langle J_f || T_J^{mag} || J_i\rangle|^2 \nonumber\\
+|\langle J_f || T_J^{elec} || J_i\rangle|^2] \nonumber\\
-2\hat{\vec{q}}\cdot(\hat{\vec{\nu}}-\vec{\beta}) Re \langle J_f || T_J^{mag} || J_i\rangle \langle J_f || T_J^{elec} ||J_i\rangle^{*}\} \nonumber\\ 
+ \sum_{J\geq0}W(E_{\nu})\{(1-\hat{\vec{\nu}}\cdot\vec{\beta}+2(\hat{\vec{\nu}}\cdot\hat{\vec{q}})(\vec{\beta}\cdot\hat{\vec{q}}))\nonumber\\
\times |\langle J_f || L_J || J_i\rangle|^2 + (1+\hat{\vec{\nu}}\cdot\vec{\beta})|\langle J_f || M_J || J_i \rangle|^2\nonumber\\
-2\hat{\vec{q}}\cdot(\hat{\vec{\nu}}+\vec{\beta})Re\langle J_f||L_J ||J_i\rangle \langle J_f || M_J|| J_i\rangle^{*}\}) , 
\end{eqnarray}
where $\vec{q} = \vec{\nu} -\vec{k}$ is the momentum transfer with $\vec{\nu}$ and $\vec{k}$ the neutrino and electron momentum, respectively, $\hat{\vec{q}}$ and $\hat{\vec{\nu}}$ are the corresponding unit vectors, and $\vec{\beta}$ =$\vec{k}/E_e$.   
$G_F$ is the Fermi coupling constant, $F(Z, E_e)$ is the Fermi function, and
$W(E_{\nu})$ is the neutrino phase space given by 
\begin{equation}
W(E_{\nu}) = \frac{E_{\nu}^2}{1+E_{\nu}/M_{T}}, 
\end{equation}
where $E_{\nu} = E_e -Q +E_i -E_f$ is the neutrino energy and $M_T$ is the target mass. The $Q$ value is determined from $Q = M_i -M_f$, where $M_i$ and $M_f$ are the masses of parent and daughter nuclei, respectively.
The Coulomb, longitudinal, transverse magnetic, and electric multipole operators with multipolarity $J$ are denoted as $M_J$, $L_J$, $T_J^{mag}$ and $T_J^{elec}$, respectively.

For a 0$^{+}$ $\rightarrow$ 2$^{+}$ transition, the transition matrices for Coulomb, longitudinal, and electric transverse operators from a weak vector current as well as an axial magnetic operator from weak axial-vector current with multipolarity $J=2$ contribute to the rates: 
\begin{eqnarray}
M_2(q)+L_2(q) = F_1^{V}(q^2)\frac{q_{\mu}^2}{q^2} j_2(qr)Y^{2}\nonumber\\
T_2^{elec}(q) = \frac{q}{M}F_1^{V}(q^2) (\sqrt{\frac{3}{5}}j_1(qr)[Y^1 \times\frac{\vec{\nabla}}{q}]^2 \nonumber\\
-\sqrt{\frac{2}{5}}j_3(qr)[Y^3 \times\frac{\vec{\nabla}}{q}]^2)
+\frac{1}{2}\mu_V(q^2) j_2(qr)[Y^2 \times\sigma]^2 \nonumber\\
T_2^{mag,5}(q) = F_A(q^2) j_2(qr)[Y^2 \times\sigma]^2, 
\end{eqnarray}
where  $F_1^{V}$, $\mu^{V}$ and $F_A$ are the nucleon vector (Dirac), magnetic, and axial-vector form factors, respectively \citep{Kura}.

Here, we evaluate the electron-capture rates for the forbidden transition, $^{20}$Ne (0$_{g.s.}^{+}$) $\rightarrow$ $^{20}$F (2$_{g.s.}^{+}$), with the USDB shell-model Hamiltonian \citep{BR} within the $sd$ shell as well as the YSOX Hamiltonian \citep{Yuan}.  The YSOX Hamiltonian, designed to be used in $p-sd$ shell configuration space, can reproduce well the ground state energies and energy levels, electric quadrupole properties, and spin properties of boron, carbon, nitrogen, and oxygen isotopes.  
Calculated e-capture rates for the forbidden transition obtained with the USDB and YSOX Hamiltonians
are shown in Fig. 1 for $\log_{10}~T$(K) $= 8.6$. 
Here, the quenching factors for the axial-vector coupling constant $g_A$ are taken to be $q =0.764$ \citep{RB} and $q=0.85$ \citep{Yuan} for the USDB and YSOX, respectively. Harmonic oscillator wave functions with a size parameter $b=1.85$ fm are used.  
Calculated rates obtained as an allowed transition with a $B$(GT) value corresponding to log{\it ft} = 10.47 \citep{Kirsebom}, that is, $B$(GT) $= 1.04 \times 10^{-6}$,  are also shown in Fig. 1.   
We refer to this method as $''$ GT prescription $''$ hereafter.     
A sizable difference is found between the two methods. The rates obtained by the GT prescription are found to be enhanced (reduced) compared with those with the USDB and YSOX at log$_{10}$($\rho Y_e$) $<$ ($>$) 9.9.

\begin{figure*}[tbh]
\begin{center}
\includegraphics[scale=1.20]{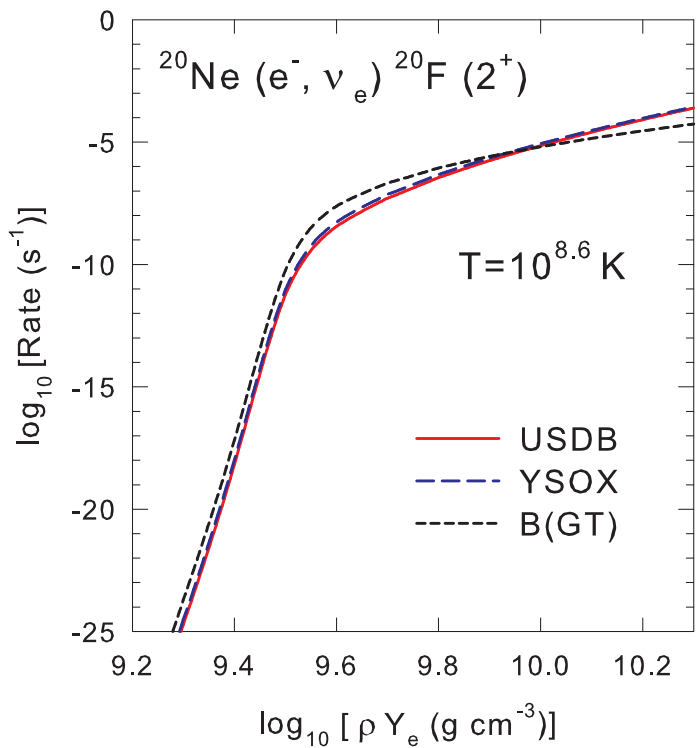}
\caption{
Calculated e-capture rates for $^{20}$Ne (e$^{-}$, $\nu_e$) $^{20}$F (2$_{g.s.}^{+}$) at $T = 10^{8.6}$ (K) obtained with the shell-model USDB \citep{BR} and YSOX \citep{Yuan} Hamiltonians as well as the GT the prescription that treats the transition as a GT one with $B$(GT) $= 1.04 \times 10^{-6}$ determined from the inverse $\beta$-decay rate of log~{\it ft} = 10.47 \citep{Kirsebom}. 
\label{fig:fig1}}
\end{center}
\end{figure*}

These tendencies are due to the difference in the electron energy dependence of the reaction cross section $\sigma (E_e)_{f,i}$ between the two methods.
$\sigma(E_e)$ for the shell-model calculations with USDB and YSOX as well as for the prescription of using the constant $B$(GT) value are shown in Fig. 2(a). 
The cross section for the constant $B$(GT) is proportional to the neutrino phase-space factor $W(E_{\nu})$, that is, it increases nearly proportional to $E_{\nu}^2$ as the electron energy $E_e$ increases. Different electron energy dependence of the cross sections is found for the shell-model results: the cross sections are reduced (enhanced) at $E_e$ $<$ ($>$) 9.9 MeV compared with the $B$(GT) prescription.
In the case of the shell-model calculations, 
the contributions from the axial magnetic and transverse electric terms are dominant at low $E_e$ regions, while those from the Coulomb and longitudinal terms increase as $E_e$ increases: 
the latter contributions become 26.9$\%$ (14.5$\%$), 48.0$\%$ (30.1$\%$), 55.7$\%$ (37.1$\%$), and 59.1$\%$ (41.3$\%$) of the total ones at $E_e$ = 9, 11, 13, and 15 MeV, respectively, for USDB (YSOX), and they finally reach almost constant fractions of 62$\%$ (44$\%$) at $E_e \geq$ 20 MeV.

\begin{figure*}[tbh]
\includegraphics[scale=0.95]{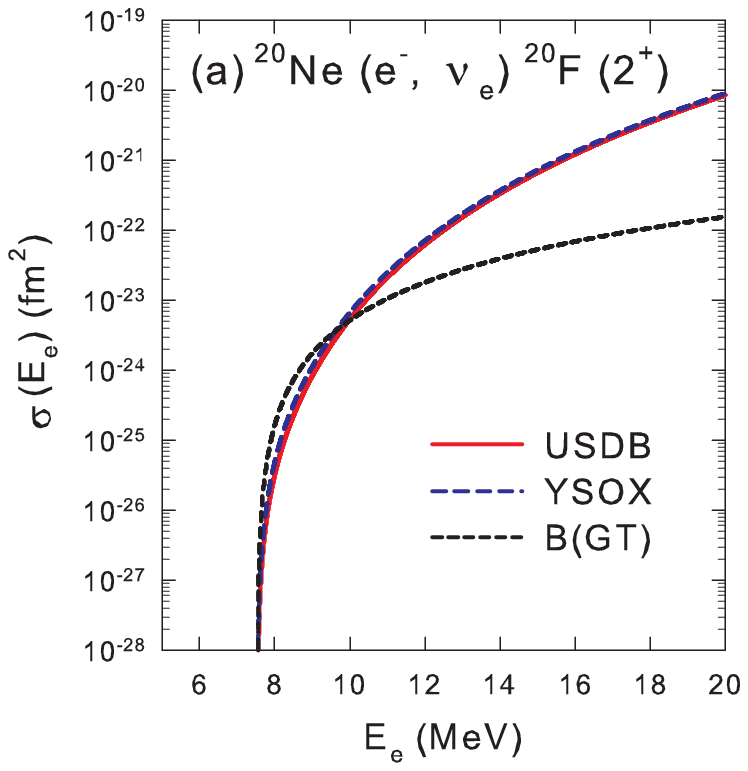}
\includegraphics[scale=0.95]{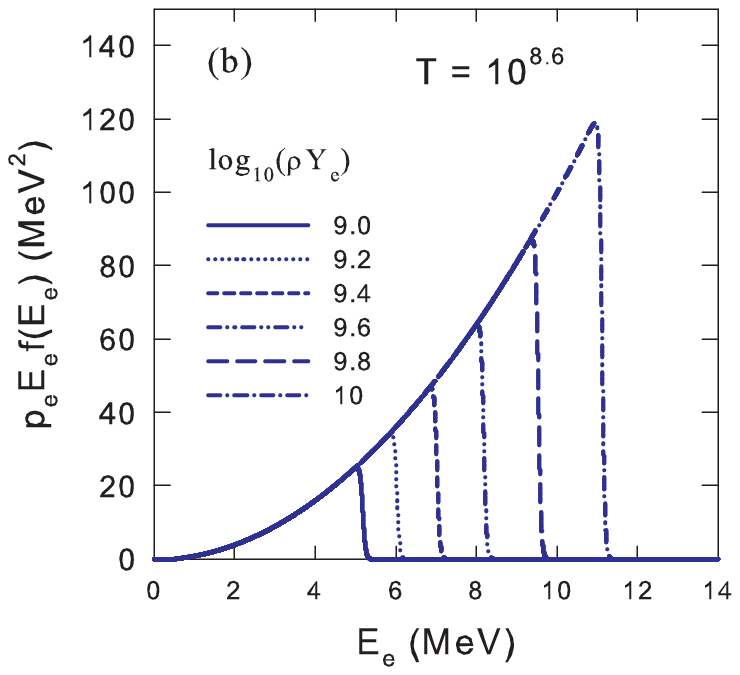}
\caption{
(a) Cross sections $\sigma (E_e)$ defined in Eq. (2) for the e-capture process $^{20}$Ne (e$^{-}$, $\nu_e$) $^{20}$F (2$_{g.s.}^{+}$) obtained with USDB, YSOX, and the prescription with the $B$(GT) value corresponding to the $\beta$-decay rate of log~{\it ft} = 10.47.
(b) Fermi-Dirac distribution of electron at $T$ = 10$^{8.6}$ (K) multiplied by the electron phase-space factor for log$_{10}(\rho Y_e) = 9.0-10.0$ in steps of 0.2.
\label{fig:fig2}}
\end{figure*}

The Fermi distribution of electron multiplied by the electron phase-space factor, $p_e E_e f(E_e)$  is shown in Fig. 2(b) for the densities $\log_{10}~(\rho Y_e$ (g cm$^{-3})) = 9.0-10.0$ in steps of 0.2.  As the density increases, the chemical potential of electron increases and the region of $E_e$ that can contribute to the e-capture rates increases.    
At log$_{10}$($\rho Y_e$) $<$ 9.8, where the electron chemical potential is below 10 MeV, the shell-model cross sections are smaller than the $B$(GT) one, and the shell-model rates also remain smaller than the $B$(GT) one. At log$_{10}$($\rho Y_e$) $\geq$ 9.9, the electron energy larger than 10 MeV can contribute to the rates, and the shell-model rates begin to exceed the $B$(GT) rate. 

Next, we study the effects of the Coulomb effects. 
Screening effects on both electrons and ions are taken into account for the Coulomb effects \citep{Juod,TSN,STN}. 
The screening effects of electrons are evaluated by using the dielectric function obtained by relativistic random phase approximation \citep{Itoh}. 
The effect is included by reducing the chemical potential of electrons by an amount equal to the modification of the Coulomb potential at the origin, $V_s(0)$ \citep{Juod}, where
\begin{eqnarray}
V_{s}(r)
&=& Ze^{2}(2k_{F})J(r) \nonumber\\
J(r) &=& \frac{1}{2 k_{F}r} \left(1- \frac{2}{\pi}\int \frac{\sin(2k_{F}qr)}{q^{2}\epsilon(q,0)}dq\right).
\end{eqnarray}
The screening coefficient $J$ is tabulated in \citet{Itoh}.

The other Coulomb effect is the change of the threshold energy 
\begin{equation}
\Delta Q_C = \mu_C(Z-1) -\mu_C(Z),
\end{equation}
where  $\mu_C(Z)$ is the Coulomb chemical potential of the nucleus with charge number $Z$ due to the interactions of the ion with other ions in the electron background
\citep{Slat,Ichimaru}.  
The Coulomb chemical potential in a plasma of electron number $n_e$ and temperature $T$ is given by
\begin{equation}
\mu_{C}(Z) = kT f(\Gamma)
\end{equation}
with $\Gamma$ = $Z^{5/3}\Gamma_e$, $\Gamma_e$ = $\frac{e^2}{kTa_e}$ and $a_e$ = $(\frac{3}{4\pi n_e})^{1/3}$.
The function $f$ for the strong-coupling regime, $\Gamma >$1, is given by Eq. (A.48) in \citet{Ichimaru}, while for the weak-coupling regime an analytic function given by Eq. (A.6) in \citet{Juod} is used for $\Gamma <$1 \citep{Yakovlev}.   
The threshold energy gets larger for e-capture processes. 
The e-capture ($\beta$-decay) rates are thus reduced (enhanced) by both the Coulomb effects.

Calculated results for the e-capture rates for the USDB Hamiltonian with and without the Coulomb effects are shown in Fig. 3(a).
The Coulomb effects shift the e-capture rates toward a higher-density region due to an increase of the $Q$ value: 
The e-capture rates with the Coulomb effects for the USDB, YSOX Hamiltonians as well as the GT prescription are shown in Fig. 3(b).
The shell-model rates are reduced compared with the GT one at 9.6 $<$ log$_{10}$($\rho Y_e$) $<$ 9.9.
The difference of the rates of the two methods is at most about by 5 times at log$_{10}$($\rho Y_e$) = 9.6.

\begin{figure*}[tbh]
\includegraphics[scale=0.95]{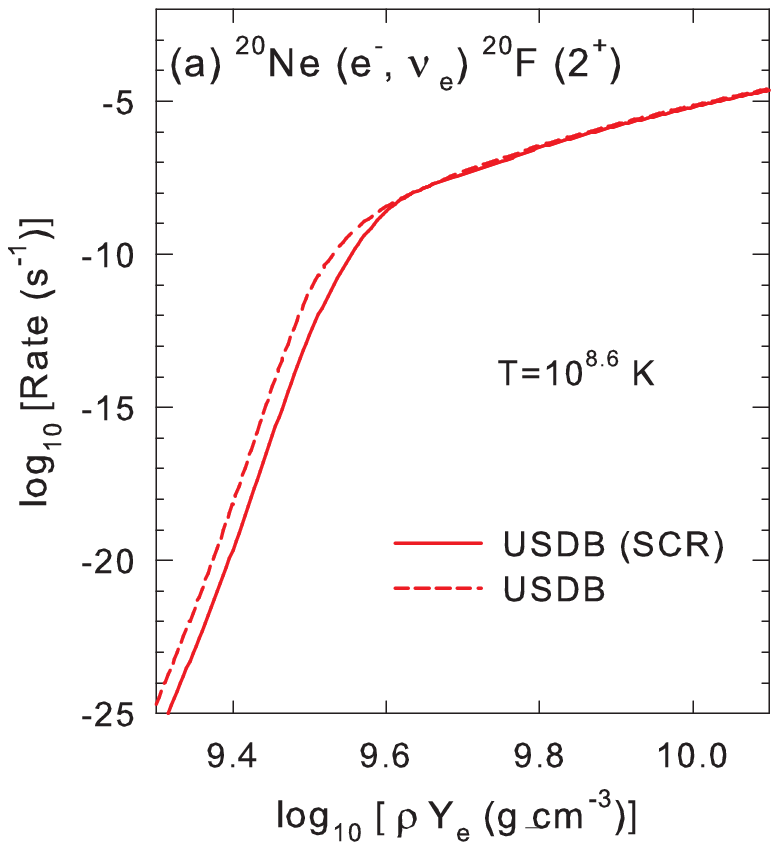}
\includegraphics[scale=0.95]{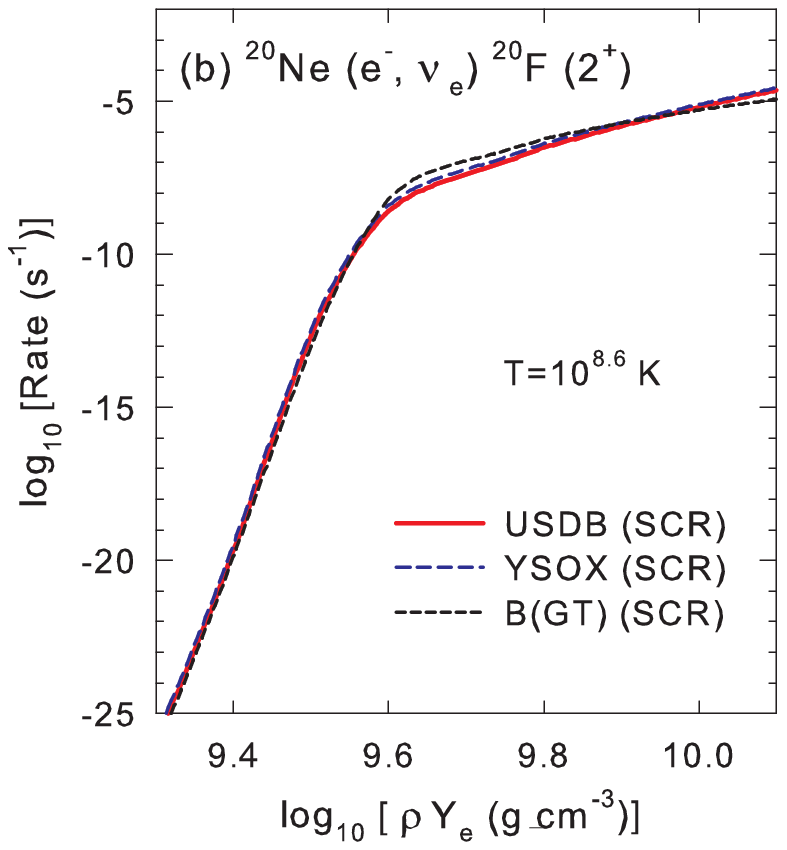}
\caption{
(a) Calculated e-capture rates for $^{20}$Ne (e$^{-}$, $\nu_e$) $^{20}$F (2$_{g.s.}^{+}$) at $T = 10^{8.6}$ (K) obtained with USDB with and without the Coulomb (SCR) effects.
(b) Same as in Fig. 1, but with the Coulomb effects (SCR). 
\label{fig:fig3}}
\end{figure*}

Total e-capture rates on $^{20}$Ne for the USDB and the GT prescription for the forbidden transition are shown in Fig. 4 for the case with the Coulomb effects.   
Contributions from Gamow-Teller transitions from 0$_{g.s.}^{+}$ and 2$_{1}^{+}$ states in $^{20}$Ne to 1$^{+}$, 2$^{+}$ and 3$^{+}$ states in $^{20}$F evaluated with the USDB are included as well as the forbidden transition, 0$_{g.s.}^{+}$ $\rightarrow$ 2$_{g.s.}^{+}$.
The difference of the two methods is at most about by 3 (3-4) times
at $\log_{10}(\rho Y_e) = 9.5 - 9.7$ for $T$ = 10$^{8.6}$ (10$^{8.4}$) (K).
In case of $T$ = 10$^{9.0}$ (K), the difference disappears.
As the calculated rates for USDB and YSOX are similar, we show results for the USDB only hereafter.

\begin{figure*}[tbh]
\begin{center}
\includegraphics[scale=0.95]{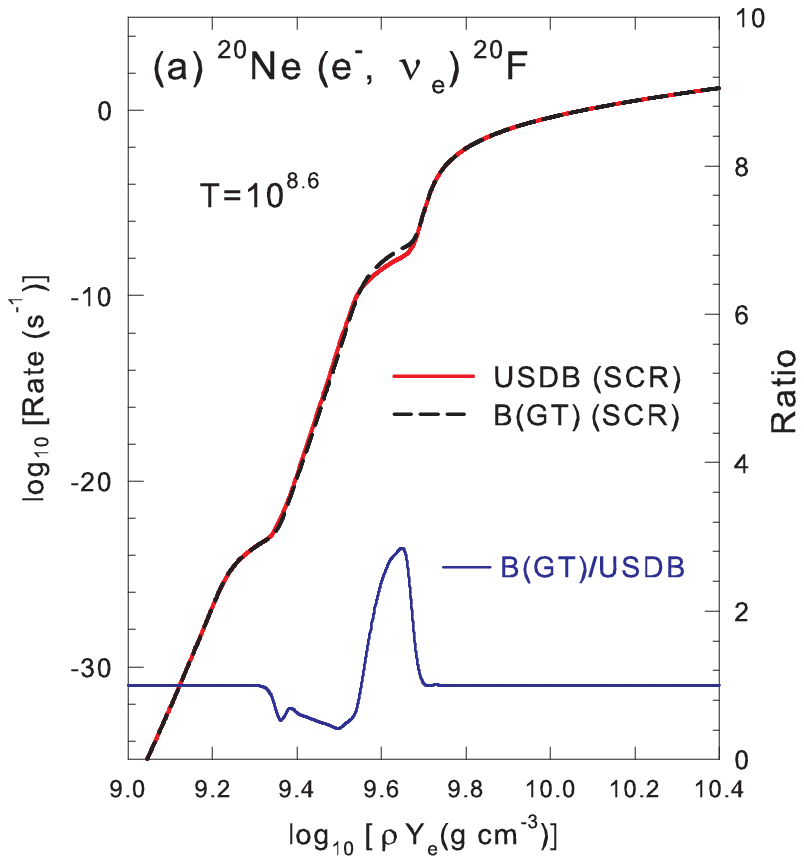}
\includegraphics[scale=0.95]{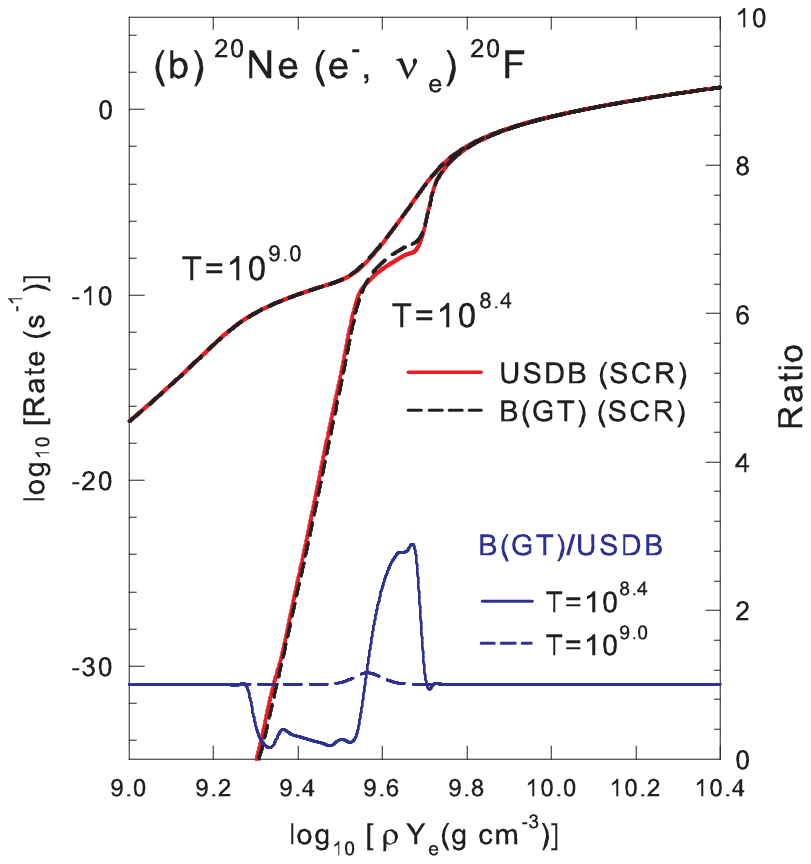}
\caption{
(a) Calculated total e-capture rates for $^{20}$Ne (e$^{-}$, $\nu_e$) $^{20}$F at $T$ = 10$^{8.6}$ (K) obtained with the Coulomb effects (SCR). The forbidden transition for $^{20}$Ne (0$_{g.s.}^{+}$) $\rightarrow$ $^{20}$F (2$_{g.s.}^{+}$) is obtained with USDB or the GT prescription.  The other GT transitions are obtained with USDB.
The ratio of the rates, B(GT)/USDB, is also shown.
(b) Same as in (a), but for $T$ = 10$^{8.4}$ and 10$^{9.0}$ (K). 
\label{fig:fig4}}
\end{center}
\end{figure*}

Now we discuss $\beta$-decay rates for the forbidden transition, $^{20}$F (2$_{g.s.}^{+}$) $\rightarrow$ $^{20}$Ne (0$_{g.s.}^{+}$). 
The $\beta$-decay rate for finite density and temperature is given as \citep{Ocon,Walecka},
\begin{eqnarray}
\lambda^{\beta}(T) = \frac{2 V_{ud}^2 g_V^2 c}{\pi^2 (\hbar c)^3}
\int_{m_e c^2}^{Q} S(E_e,T) E_e p_e c (Q-E_e)^2\nonumber\\
\times (1 -f(E_e)) dE_e \nonumber\\
S(E_e,T) = \sum_{i} \frac{(2J_i +1)e^{-E_i/kT}}{G(Z,A,T)}
 \sum_{f} S_{f,i}(E_e)\nonumber\\
S_{f,i}(E_e) = \int \frac{1}{4\pi} d\Omega_{\nu} \int d\Omega_k
\frac{G_F^2} {2\pi} \frac{F(Z+1,E_e)}{(2J_i+1)}\nonumber\\
\times (\sum_{J\leq1}
\{(1-(\hat{\vec{\nu}}\cdot\hat{\vec{q}})(\vec{\beta}\cdot\hat{\vec{q}}))  
[|\langle J_f || T_J^{mag} || J_i\rangle|^2\nonumber\\
+|\langle J_f || T_J^{elec} || J_i\rangle|^2]\nonumber\\
+2\hat{\vec{q}}\cdot(\hat{\vec{\nu}}-\vec{\beta})
Re \langle J_f || T_J^{mag} || J_i\rangle \langle J_f || T_J^{elec} ||J_i\rangle^{*}\} \nonumber\\ 
+ \sum_{J\geq0}\{(1-\hat{\vec{\nu}}\cdot\vec{\beta}+2(\hat{\vec{\nu}}\cdot\hat{\vec{q}})(\vec{\beta}\cdot\hat{\vec{q}}))
|\langle J_F || L_J || J_i\rangle|^2\nonumber\\ 
+ (1+\hat{\vec{\nu}}\cdot\vec{\beta})|\langle J_f || M_J || J_i \rangle|^2 \nonumber\\
-2\hat{\vec{q}}\cdot(\hat{\vec{\nu}}+\vec{\beta})Re\langle J_f||L_J ||J_i\rangle \langle J_f || M_J|| J_i\rangle^{*}\}),  
\end{eqnarray}
where $\vec{q}$ = $\vec{k}+\vec{\nu}$, and the factor $1-f(E_e)$ denotes the blocking of the decay by electrons in high-density matter.

The log~{\it ft} value for a $\beta$-decay transition is given as \citep{Oda,Lan1} 
\begin{eqnarray}
ft &=& {\rm ln} 2 \frac{I}{\lambda^{\beta}} \nonumber\\ 
I &=& \int_{m_e c^2}^{Q} E_e p_e c (Q-E_e)^2 F(Z+1, E_e)\nonumber\\
& &\times (1 -f(E_e)) dE_e. 
\end{eqnarray}
Here, $\lambda^{\beta}$ is the $\beta$-decay rate for the transition, and $I$ is the phase-space integral. 
In the case of $\beta$-decay in the vacuum at $T$ = 0 or in low-density matter at low temperature, the term $(1 -f(E_e)$) can be replaced by 1.

The transition strengths multiplied by phase-space factors are compared in Fig. 5 for USDB and the GT prescription. 
A small dent seen around $E_e$ = $\frac{1}{2}Q$ for the GT prescription arises from the lepton kinematical factor for the transverse multipoles, $f_T$ =      $1-(\hat{\vec{\nu}}\cdot \hat{\vec{q}})(\vec{\beta}\cdot\hat{\vec{q}})$. 
It can be expressed as 
$f_T$ = $(1 -\beta\frac{\omega^2}{\vec{q}^2}) cos^2(\theta/2) + (1+\beta) sin^2(\theta/2)$ with cos($\theta$) = $\hat{\vec{k}}\cdot\hat{\vec{\nu}}$. 
Here $\omega$ =$k+\nu$ ($k=|\vec{k}|$ and $\nu =|\vec{\nu}|$) is the energy transfer, and $\beta$ =$|\vec{\beta}|$ =$k/E_e$.
An integral $f$=$\int_{0}^{\pi} f_T 2\pi sin\theta d\theta$ is a function of $E_e$ with a minimum at $E_e$ =$Q$/2 and maxima at $E_e$ = $m_e$ and $Q$, where $m_e$ is the electron mass, 
leading to a divot at $E_e =\frac{1}{2}Q$.
However, this behavior of the strength little affects the $\beta$-decay transition rate. 
The difference in the decay rate (log~{\it ft} value) from that obtained by a standard formula for allowed $\beta$-decay, where $f$ is taken to be a constant, is as small as 15$\%$ (0.06).    
 
The strength of USDB is reduced compared with the GT prescription in the whole energy region.
The summed strength is also small for the shell-model case, and leads to a log~{\it ft} value for USDB larger by 0.65 compared with the GT prescription, that is, log~{\it ft} = 11.18. The branching ratio is obtained to be 2.15$\times 10^{-6}$, which is 19.5$\%$ of the observed value \citep{Kirsebom}.
Sensitivity to radial behavior of wave functions is examined by using Woods-Saxon wave functions obtained with standard parameters of Ref. \citep{BM}.
The difference from the harmonic oscillator case is rather small except at $E_e <$ 2 MeV as shown in Fig. 5.
The summed strength is increased only by 5$\%$.  
We also give calculated results for YSOX; log~{\it ft} =11.24 and the branching ratio is 1.87$\times 10^{-6}$.
The difference of the strengths between the two methods can be ascribed to that in the energy dependence of the strengths, which prove to be important in explaining the difference in the e-capture rates.

\begin{figure*}[tbh]
\begin{center}
\includegraphics[scale=1.2]{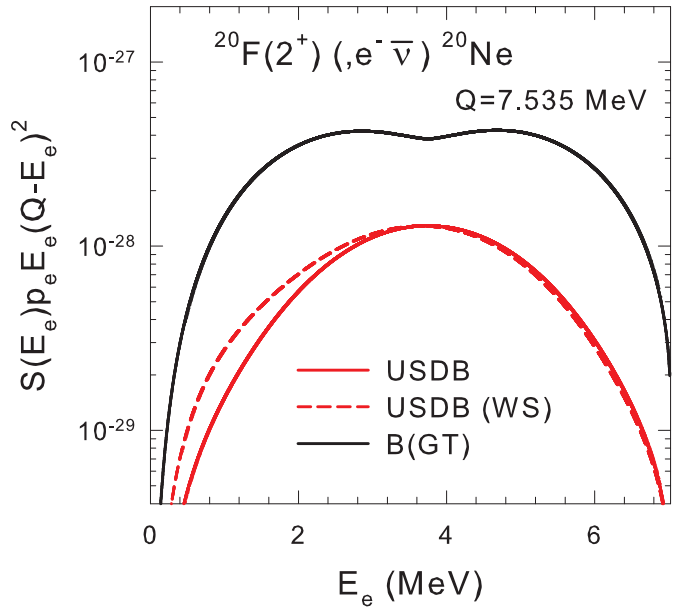}
\caption{
Transition strength S$(E_e)$ with phase space-factors defined in Eq. (8) for the $\beta$-decay process 
$^{20}$F (2$_{g.s.}^{+}$) (, e$^{-}$ $\bar{\nu_e}$) $^{20}$Ne (0$_{g.s.}^{+}$) obtained with USDB and GT prescription.
The strength for USDB obtained with Woods-Saxon wave functions 
is also shown.
\label{fig:fig5}}
\end{center}
\end{figure*}

\section{Evolution of High-density O-Ne-Mg Cores\label{sec:onemg_core}}

Now, we study the effects of the forbidden transition in the electron-capture processes on $^{20}$Ne on the evolution of the high-density electron-degenerate O$-$Ne$-$Mg cores. 
Heating of the core due to $\gamma$ emissions succeeding the double e-capture reactions, $^{20}$Ne (e$^{-}$, $\nu_e$) $^{20}$F (e$^{-}$, $\nu_e$) $^{20}$O, is important in the final stage of the evolution of the core. 
Four transition rates obtained with (1) USDB with the Coulomb effects, (2) the GT prescription with the Coulomb effects, (3) USDB without the Coulomb effects, and (4) the GT prescription without the Coulomb effects are used for the forbidden transition, $^{20}$Ne (0$_{g.s.}^{+}$) $\rightarrow$ $^{20}$F (2$_{g.s.}^{+}$), and the sensitivities of the evolution of the core on the rates are investigated.

We calculate the growth of an O$-$Ne$-$Mg core toward the Chandrasekhar mass limit using Modules for Experiments in Stellar Astrophysics \citep[MESA;]{MESA1,MESA2,MESA3}, revision 8118. The O$-$Ne$-$Mg core is prepared by evolving an $8.4~M_\odot$ nonrotating solar-metallicity star as in \citet{Jones}, removing its envelope right before thermal pulses of He shell burning start.
We increase the O$-$Ne$-$Mg core mass at a constant rate of $10^{-6}~M_\odot~yr^{-1}$, similar to \citet{Schwab19}. URCA and electron-capture processes are included using the rates in \citet{Tab}. 
For convective stability, the Ledoux criterion is adopted. We switch off the mixing-length treatment for convection by the control \texttt{mlt\_option = `none'} when electron capture on $^{24}$Na starts, as done in \citet{Schwab2017} to avoid numerical nonconvergence. The evolution with the Schwarzschild criterion needs more computationally efforts to follow, and the results will be shown elsewhere \citep{Zha}.

\begin{figure*}[tbh]
\begin{center}
\includegraphics[scale=0.50]{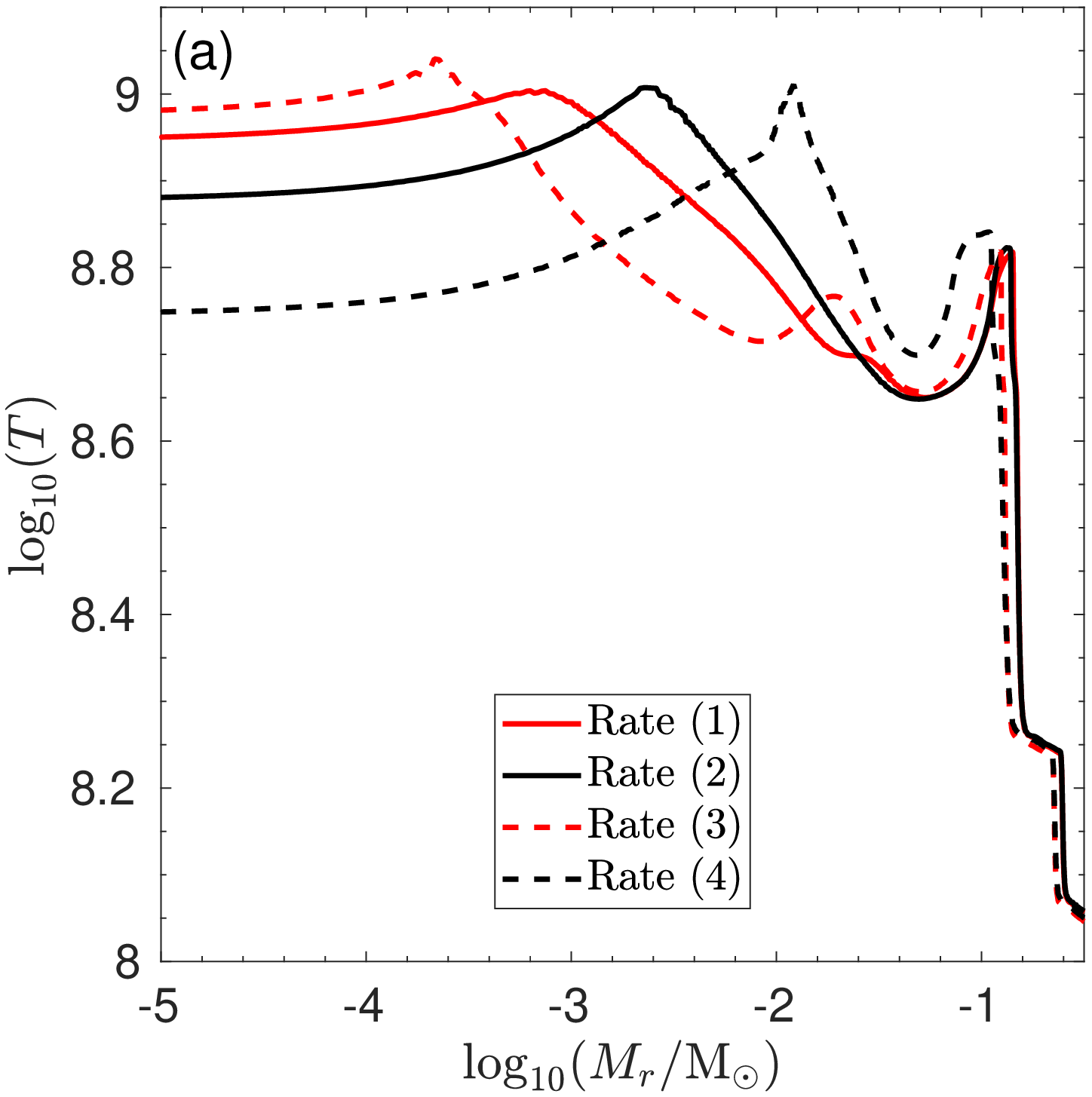}
\hspace*{0.5cm}
\includegraphics[scale=0.50]{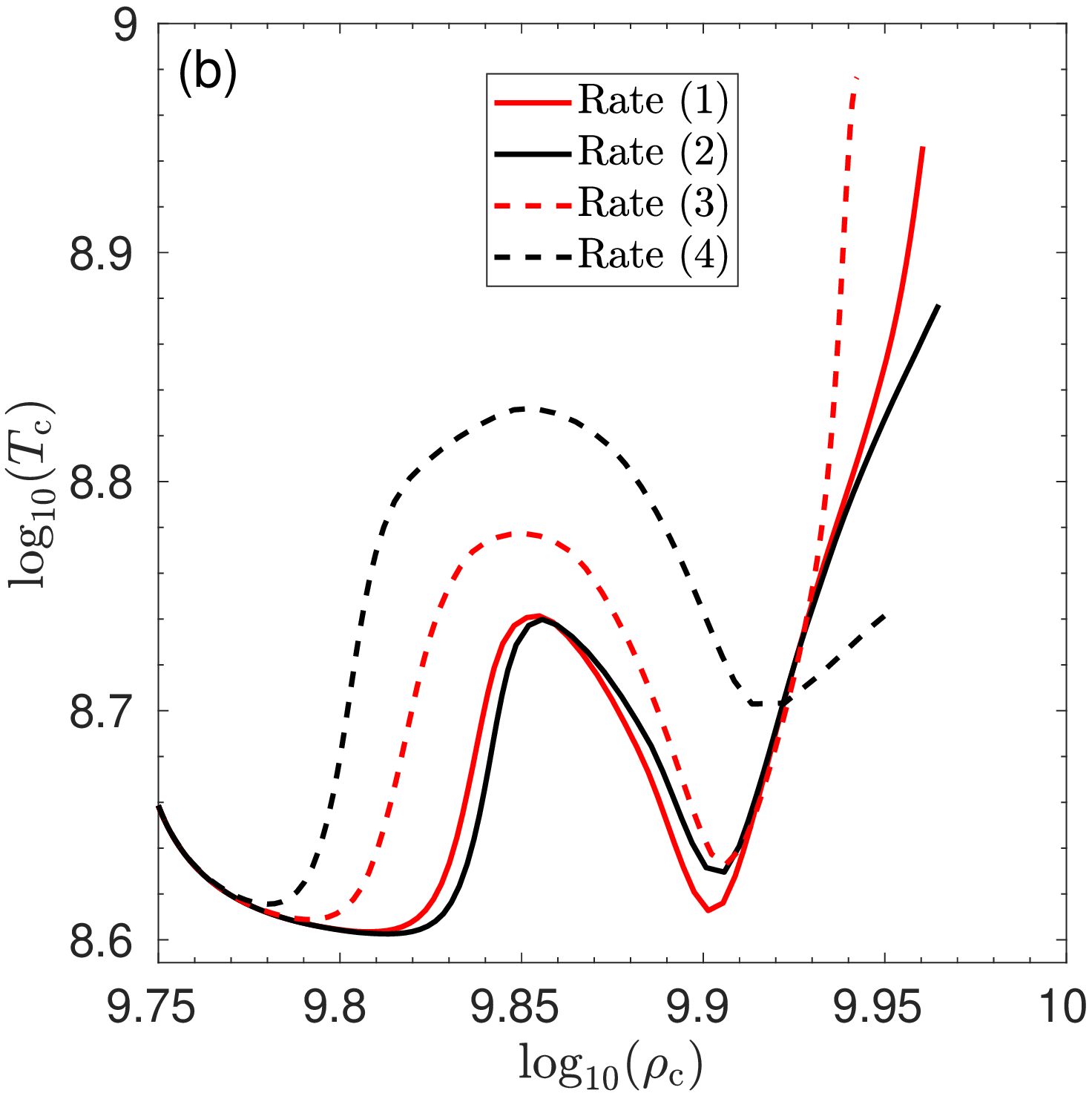}
\caption{(a) The temperature profiles as a function of the enclosed mass
 ($M_r$) at the moment of the oxygen ignition for the four different
rates described in the text. Rate (1) USDB with the Coulomb effects, (2) the GT prescription with the Coulomb effects, (3) USDB without the Coulomb effects, and (4) the GT prescription without the Coulomb effects.
(b) The evolution of central temperature ($T_{\rm c}$) and
 density ($\rho_{\rm c}$) of an O$-$Ne$-$Mg core, starting from the end
 of $^{24}$Mg$(e^{-},\nu_e)^{24}$Na$(e^{-},\nu_e)^{24}$Ne up to
 the ignition of oxygen.
\label{fig:core_evol}}
\end{center}
\end{figure*}

In Fig.~\ref{fig:core_evol}(a), 
we show the temperature distribution as a function of the enclosed mass $M_r$ for the four different rates (1)$-$(4) defined above.  
The evolution of the central density and temperature of the O$-$Ne$-$Mg core is shown in Fig.~\ref{fig:core_evol}(b) 
for the four rates.
In Fig.~\ref{fig:core_evol}(a), 
the temperature inversion appears in
the central region because of the following reason.  The electron-capture rate 
on $^{20}$Ne with the second forbidden transition
is not high enough to cause a rapid heating of the center, and then 
the central region is cooled down by the $^{25}$Na-$^{25}$Ne Urca shell cooling around log~$\rho_{\rm c} = 9.85 - 9.9$ as seen in Fig.~\ref{fig:core_evol}(b).
At log~$\rho_{\rm c} > 9.9$, the temperatures in these layers increase due to e-capture on $^{20}$Ne as well as core contraction. 
The heating effect of e-capture on $^{20}$Ne is slightly higher in the outer
layers because the outer layer contains the larger mass fraction of $^{20}$Ne 
than the inner layer when their densities reach around log~$\rho$ =9.9 during the contraction.
This leads to the formation of the temperature inversion. 
Eventually oxygen is ignited in the outer shell as seen in Fig.~\ref{fig:core_evol}(a).
The off-center oxygen ignition occurs at $6.9\times10^{-4}~M_\odot$
($33$ km), $2.3 \times10^{-3}~M_\odot$ ($50$ km),
$2.3 \times 10^{-4}~M_\odot$ ($23$ km) and $1.2 \times
10^{-2}~M_\odot$ ($88$ km) for the four different rates, respectively.  
Here the ignition occurs when the nuclear energy generation rate exceeds the thermal neutrino loss.
The heating due to oxygen ignition forms a convectively unstable region 
even for the Ledoux criterion; the resulting convective energy transport will slow down the increase in the temperature due to oxygen burning.

Further contraction of the core to the higher central density will continue before the thermonuclear runaway.  
The final central density and position of oxygen ignition are important for the subsequent hydrodynamical behavior and flame propagation, and thus the fate of the O$-$Ne$-$Mg core \citep{Jones16,Leung19,Takahashi19}. The ignition closer to the center for the USDB rates with Coulomb effects may favor neutron star formation as the final outcome \citep{Nomoto2017,Leung2018}. The detailed study of the complete evolution with these new rates, as well as how the theoretical uncertainties affect the evolutionary path, will be reported elsewhere \citep{Zha}.


\section{Summary}

We evaluated e-capture 
rates for the forbidden transition, $^{20}$Ne (0$_{g.s.}^{+}$) $\rightarrow$ $^{20}$F (2$_{g.s.}^{+}$), by the multipole expansion method of \citet{Ocon} and \citet{Walecka}. 
The Coulomb, longitudinal, transverse electric, and axial magnetic multipoles with $J^{\pi}$ = 2$^{+}$ contribute to the transitions. 
The e-capture rates at stellar environments obtained by the multipole method with the USDB and YSOX Hamiltonians are compared with the GT prescription that treats the transitions as allowed Gamow-Teller transitions with the $B$(GT) value determined from the $\beta$-decay log~{\it ft} value.
Sizable differences are found between the rates obtained by the two methods for both cases with and without the Coulomb effects.
Origin and density dependence of these differences in the rates are shown to be explained by the difference in the electron energy dependence of the transition strengths. 

The four e-capture rates on $^{20}$Ne obtained by the multipole method with the USDB and the GT prescription, with and without the Coulomb effects, are used to study the evolution of the high-density O$-$Ne$-$Mg cores.
Sizable sensitivity of the heating process in the O$-$Ne$-$Mg core on the e-capture rates is found in the final stage of the evolution of the core.
It is thus important to evaluate the weak rates of forbidden transitions properly with the multipole expansion method, which gives rise to energy dependence of the transition strengths.

\acknowledgements

The authors thank MESA developers for making their code open-source. S. Z.  acknowledge the CUHK Global Scholarship for Research Excellence for supporting his stay at Kavli IPMU during which the stellar evolution calculations were done.
This work has been supported in part by Grants-in-Aid for Scientific
Research
(C) JP19K03855, JP16H02168, JP17K05382,  
and the World Premier International Research Center Initiative
of the MEXT of Japan.
We acknowledge the support by the Endowed Research
Unit (Dark Side of the Universe) by Hamamatsu Photonics K.K.

\end{document}